\documentclass[%
 reprint,
 amsmath,amssymb,
 aps,
]{revtex4-2}

\usepackage{graphicx}
\usepackage{latexsym}
\usepackage{amsmath}
\usepackage{amssymb}
\usepackage{dblfloatfix}
\usepackage{amsfonts}
\usepackage{xcolor}
\usepackage{cases}
\usepackage[normalem]{ulem}

\begin{document}

\preprint{APS/123-QED}

\title{Solitons and vortices in non-equilibrium photon condensates}

\author{Michiel Yzewyn}
\author{Michiel Wouters}%
\affiliation{%
 TQC, Universiteit Antwerpen, Universiteitsplein 1, B-2610 Antwerpen, Belgium
}%

\date{\today}

\begin{abstract}

We investigate the stability and size of solitons and vortices in Bose-Einstein condensates of photons in dye filled microcavities using a classical field model. In contrast to interacting bosonic systems where the healing length is dictated by the interaction strength, we show analytically that the healing length for non-interacting driven-dissipative photon condensates depends on the losses and thermalization rate. We find that the solitons are not stable with respect to the spontaneous emission noise.
Vortices do persist in the presence of noise and our analytical formula for the healing length agrees well with the vortex core size, both for lattices and in the continuum limit

\end{abstract}

\maketitle


 \section{Introduction}

Nonlinear excitations, such as solitons and vortices, are a hallmark of quantum fluids. Although traditionally investigated in near-equilibrium systems, such as ultracold atomic Bose–Einstein condensates (BECs) \cite{BECbook}, these excitations have also been demonstrated in nonequilibrium quantum fluids, particularly exciton-polariton condensates \cite{Lagoudakis:NatPhys2008,carusotto:2013,bloch2022non}. 
Photon condensates in dye filled cavities \cite{klaers:NatPhys2010} form a complementary platform for the study of nonequilibrium quantum fluids.
Since photons have negligible interaction, a naive estimate of the vortex or soliton core size in analogy with equilibrium condensates diverges, but numerical simulations for photon condensates in a periodic potential have shown that the nonlinearity from the coupling to the gain reservoir results in a finite core size \cite{gladilin2020vortices}. 

In one-dimensional Bose-Einstein condensates localized excitations come under the form of solitons. In equilibrium they constitute a one-parameter family of solutions for which the depth, phase jump and velocity are related. In resonantly driven polariton condensates, solitons have been observed in the wake of an obstacle immersed in the superfluid flow \cite{amo2011polariton}. Under incoherent excitation, dark soliton solutions have been theoretically predicted to feature interesting behavior such as temporal oscillations and an abrupt decay after stable long-time evolution \cite{Xue2014Creation}. Moreover, solitons are also predicted for an off-resonant spatially localized pump  \cite{Ostrovskaya2012Dissipative} in two-dimensionsal systems where the polariton condensate has been shown to generate a self-defocusing dissipative soliton, even in the absence of a trapping potential, similar to the anti-soliton solutions in the generalized complex Ginzburg-Landau equation \cite{Aranson:RMP2002}. A theoretical study of the corresponding soliton solutions in photon condensates in dye filled microcavities is however still missing.

For what concerns the vortices in nonequilibrium polariton condensates, they have been experimentally observed to be generated by and pinned to defects \cite{Lagoudakis_2008}, generated by a localized incoherent pump pulse in a ring-shaped geometry \cite{Yao_persistent_2025}  or through a rotating potential \cite{ExPolRotatingBucket}.
Although theoretical studies have been made on the size and generation of vortices in two-dimensional photon condensates \cite{gladilin2020vortices,Dhar_2021,krauß2026regularizationvortexcoresize}, experimental confirmation of their existence is still missing.
In analogy with polariton condensates, their theoretically predicted properties include a spiral structure of the phase around the core that leads to repulsive forces between vortices and anti-vortices \cite{Aranson:RMP2002,wachtel2016electrodynamic,gladilin2017interaction}.

In this article, we address the theoretical study of solitons and complement previous studies of vortices in photon condensates.  In particular, we study their stability and derive an analytical expression for their core size.
Even though solitons are shown to be dynamically unstable, the mean-field results proves to be useful for numerically determining the healing length of the condensate. Vortices are found to be stable and highly robust to fluctuations, provided that the vortex core size is on the order of at least a few cavity sites.

To simulate driven-dissipative open quantum platforms theoretically, a computationally efficient approach for a high number of photons distributed over many spatial modes is required.
Microscopic descriptions rooted in quantum master equations offer the most rigorous framework \cite{kirton-keeling-13}, yet their huge Hilbert space dimension makes them intractable for spatially extended cavities without severe simplifications. Conversely, semiclassical Boltzmann rate equations yield a lightweight alternative  \cite{schmitt2018,walker2019collective,loirette2023photon}, though they discard phase degree-of-freedom dynamics altogether and fail to account for topological excitations and for critical phenomena such as Kardar-Parisi-Zhang (KPZ) scaling. A powerful intermediate strategy couples stochastic classical fields for the photonic component with rate equations tracking excited molecular states \cite{gladilin2020classical}—a scheme closely related to truncated Wigner approaches in microcavity polariton research \cite{Wouters:PRB2009}. In this picture, Hamiltonian evolution governs phase gradient-driven supercurrents, while non-Hermitian terms enforce gain and decay balance.
Dye-filled cavities offer a distinct modeling advantage over polariton condensates: the emission and absorption properties of organic dyes are experimentally very well characterized. This precise knowledge of the dye species allows one to construct quantitative first-principles models for photon condensation with minimal free parameters \cite{klaers2010bose,sazhin2024observation}.


A full analytical solution of our classical field model is elusive, but we show that the soliton and vortex core sizes can be derived analytically from the analysis of linear excitations, both for the case of a cavity array and for a continuous photon condensate. We find a very good agreement between the soliton width and our analytically derived healing length. However, simulations that include the fluctuations stemming from spontaneous emission show that the soliton solution in photon condensates is not stable. 
Also the vortex core size in two-dimensional photon condensates is determined by the same healing length. Depending on the parameters, the core structure varies significantly, showing strong asymmetries in certain regimes, but for the majority of parameter space the healing length determines the spatial extent of the core. Their better topological protection implies a greater stability for vortices compared to solitons in the presence of noise. In finite systems, vortices do not persist indefinitely because they can leave the system under their stochastic dynamics. We find that their life time increases roughly exponentially with system size and is already on the order of nanoseconds for moderate system sizes. 



The article is organized as follows. In Sec. \ref{sec:model} the stochastic classical field model is introduced. From this model a healing length is derived in Sec. \ref{sec:analytical}, which is compared to the healing length determined from the mean-field soliton simulations in Sec. \ref{sec:soliton}. Sec. \ref{sec:vortex} shows the healing length and stability of vortices for mean-field simulations. Additionally, we investigate the scaling of the vortex life time with system size for various parameter regimes and also the continuum limit of the mean-field model is verified to be well defined. Our conclusions and outlook are given in \ref{sec:conclusion}.

\section{Model}\label{sec:model}

We will use for our theoretical analysis of solitons and vortices in photon condensates a
classical stochastic description of the site and time dependent photon amplitude $\psi(x,t)$, \cite{gladilin2020classical}. For a one dimensional array of coupled cavities the equation is given by 
\begin{align}
i \frac{\partial \psi(x,t) }{\partial t}  = 
&- (1-i\kappa) J  \left[ \psi(x-1,t) +\psi(x+1,t) - 2 \psi(x,t) \right] \nonumber \\
&+
\frac{i}{2}\left[ B_{21}M_2(x,t) - B_{12}
M_1(x,t)-\gamma\right]\psi(x,t) \nonumber \\
&+ \sqrt{D(x,t)} \; \xi(x,t).
 \label{eq:gGP}
\end{align}
Here, $M_1(x,t)$ and $M_2(x,t)$ denote the number of ground state and excited dye molecules per cavity, respectively. The parameter $J$ represents the tunneling rate between neighboring cavities and $\gamma$ is the cavity loss rate.  The Einstein absorption (emission) coefficients $B_{12}\, (B_{21})$ of the dye at the frequency of a single cavity are connected via the Kennard-Stepanov relation \cite{kennard,stepanov,moroshkin}. The Kennard-Stepanov relation introduces the dimensionless parameter $\kappa$, which is responsible for the imprinting of the temperature of the molecular bath onto the photon field.
\begin{equation}
    \kappa = \frac{B_{12} M_1 }{2 k_B T},
\end{equation}
with $k_B$ the Boltzmann constant. The stochastic nature of Eq. \eqref{eq:gGP} stems from the last term $\xi$, which has zero mean and the autocorrelation function $\langle \xi^*(x,t) \xi(x',t') \rangle  =  \delta_{x,x'} \delta(t-t')$. The noise is introduced by the random phase of the photons from the dissipative processes, with corresponding magnitude $D(x,t) = (B_{12} M_1(x,t) + B_{21} M_2(x,t) + \gamma)/2$. In the absence of losses, the fluctuation-dissipation relation is satisfied by the noise strength with the friction parameter $\kappa$. 
For numerical simplification, we replace the time-dependent noise with the time-independent one $D=B_{12} M+\gamma/2$, making the noise term in Eq. \eqref{eq:gGP} additive, which we have verified numerically to have a negligible effect on the photon statistics. Furthermore, this is valid in experimental regimes, where the relative population inversion $(M_2-M_1)/M$, with $M$ the total number of molecules, is much smaller than one. The terms on the r.h.s. of Eq. \eqref{eq:gGP} that are proportional to $\kappa$ correspond to the model A relaxational dynamics that drive the system into the thermal equilibrium state \cite{hohenberg1977theory}. This results in a  steady state that is inherently nonequilibrium,  since the  real and imaginary parts of the terms on the r.h.s. of Eq. \eqref{eq:gGP} do not match,  \cite{Sieberer_2013}. 

Coupled to the equation for the photon field is a rate equation for the number of excitations $X = n + M_2$ :
\begin{equation}
\frac{\partial }{\partial t} X(x,t) = - 2J\Im[\psi^*(x,t)\frac{\partial}{\partial x}\psi(x,t)] - \gamma  |\psi(x,t)|^2 + P ,
\label{eq:rate}
\end{equation}
where the first term on the r.h.s. represents the tunneling photon currents. The mirror losses are represented by the second term, with $n(x,t) = |\psi(x,t)|^2$ denoting the local photon number, which is balanced by pump $P=\gamma \bar n$, where $\bar n$ is the average photon number.
 
By looking at the linear dynamics of the photon number fluctuations $\delta n=n-\bar n$ without losses, we can derive the number fluctuation decay rate $\Gamma$ as 
\begin{equation}
\Gamma = \left(1+\frac{n^2}{M_{\rm eff}}\right)\frac{B_{12} M}{\bar n},
\label{eq:Gamma}
\end{equation}
where, the effective reservoir size is given by
\begin{equation}
M_{\rm eff} = \frac{ M}{2[1+\cosh(\beta\Delta)]}.
\end{equation} 
The simplified stochastic model for the photon deviation is found as
\begin{equation}
\frac{d}{dt} \delta n = -\Gamma\, \delta n + \sqrt{D_n}\, \xi_n.
\label{eq:deltanhom}
\end{equation}
The term $\xi_n$ respresents real Gaussian noise and has the autocorrelation function $\langle \xi_n(x,t) \xi_n(x',t') \rangle  =  \delta_{x,x'} \delta(t-t')$ and magnitude $D_n = 2 B_{12} M \bar n$. Using Eq. \eqref{eq:deltanhom} to compute the density fluctuations, \cite{verstraelen2019temporal}, results in
\begin{equation}
\langle \delta n^2 \rangle = \frac{\bar n^2}{1+\frac{\bar n^2}{M_{\rm eff}}}.
\label{eq:densfluct}
\end{equation}
In the limit of $M_{\rm eff} \gg \bar n^2$, the photon condensate is in the grandcanonical regime with large density fluctuations, while for  $M_{\rm eff} \ll \bar n^2$, the condensate is in the canonical regime with small density fluctuations, \cite{schmitt2014observation}. 
Dye-photon condensates can therefore both exhibit laser  like ($g^{(2)}=1$) and thermal light like ($g^{(2)}=2$) density fluctuations \cite{schmitt2014observation}, in contrast to polariton condensates where the statistics above condensation threshold is observed to be laser like \cite{adiyatullin2015temporally,kim2016coherent}.
As shown in Ref. \cite{loirette2023photon}, the elevated number fluctuations in photon condensates stem primarily from weak gain saturation rather than the thermalization mechanism itself.


\section{Analytical derivation of Healing length \label{sec:analytical}}


For isolated bosonic systems, the characteristic length scale for their response to localized spatial perturbations is given by the healing length $\xi=1/\sqrt{m g n}$, where $m$ is the mass, $n$ the density and $g$ the interaction constant. In the limit of vanishing interactions which is relevant for photonic condensates, the healing length diverges. In the case of driven-dissipative systems however, the nonlinear coupling between the condensate and the reservoir gives another mechanism for the appearance of a finite healing length.

This length scale  can in first approximation be determined by considering the limiting case of small perturbations on top of a homogeneous condensate steady state background.
The dynamics can then be simplified by linearizing the equations of motion in momentum space. The condensate field is therefore written as
\begin{equation}
  \psi(x,t) = \sqrt{\bar n+\delta n(x,t)} e^{i \delta \theta(x,t)},
\end{equation}
where $\delta n$ and $\delta \theta$ are small density and phase fluctuations respectively, that are decomposed in a Fourier series:
\begin{align}
    \delta n(x,t) &= \sum_k \delta n_k(t) e^{i k x},  \\
     \delta \theta(x,t) &= \sum_k \delta \theta_k(t) e^{i k x}.
\end{align}
These photon density and phase fluctuations are coupled to the fluctuation of the total number of excitations
\begin{align}
    X(x,t) = X_0 + \sum_k \delta X_k(t) e^{i k x}.
\end{align}
The linearized equations of motion can then be written in terms of the fluctuation and force vectors,
\begin{equation}
    \delta \mathbf{\Psi}_k  = \begin{pmatrix}\delta \theta_k \\\delta n_{k} \\\delta X_k\end{pmatrix} \hspace{0.25cm} {\rm and} \hspace{0.25 cm}
   \mathcal{F}_k =\begin{pmatrix}F_{\theta} \\F_{n} \\ F_X\end{pmatrix},
\end{equation}
as 
\begin{align}
    i\frac{\partial}{\partial  t} \delta \mathbf{\Psi}_k & =B_k \delta \mathbf{\Psi}_k  + \mathcal{F}_k e^{-i\omega t}, 
\end{align}
where the matrix $B_k$ is given by
\begin{align}
    B_k = \begin{bmatrix}
    -\kappa\epsilon_k & -\frac{\epsilon_k}{2n} & 0 \\
  2 n\epsilon_k & -(\Gamma+\kappa\epsilon_k)& \frac{\sigma_n^2 \Gamma}{M_{eff}} \\
 2 n\epsilon_k & -\gamma  &0\end{bmatrix}.
\end{align}


For a time-independent forcing, the solutions read
\begin{align*}
    \mathbf{\delta \Psi_k} = -B^{-1}_k \mathcal{F}_k,
\end{align*}
from which the real space perturbation can be determined by means of the inverse Fourier transform:
\begin{align}
    \delta \boldsymbol{\Psi}(x) = - \int e^{ikx} B^{-1}_k \mathcal{F}_k \; dk.   
\end{align}
Performing this integral as a contour integral, the contributions come from the poles of $B^{-1}_k$ that are the zeros of $\det(B_k)$ as a function of $k$. These momenta determine the inverse of the characteristic length scales at which the system responds to stationary spatially localized external perturbations.
They are given by $k=0$ (corresponding to the global phase rotation due to the spontaneous U(1) symmetry breaking) and by $\epsilon_k = -\gamma \kappa$.

In the long-wavelength limit $\epsilon_k = 2J(1-\cos{(k)})\approx Jk^2$, the imaginary roots of $k$ are found as $k = \pm i \sqrt{\frac{\gamma \kappa}{J}}$. This implies the healing length 
\begin{align} \label{eq:healinglength}
     \xi & = \sqrt{\frac{J}{\gamma \kappa}},
\end{align}
which agrees with the $\xi \propto \sqrt{J/\gamma}$ relation found in previous work from dimensional analysis \cite{gladilin2020vortices}.

As we will show in the next sections, this value of the healing length corresponds well to the spatial sizes of soliton and vortex solutions obtained from the exact dynamics from numerically simulating the coupled nonlinear equations \eqref{eq:gGP} and \eqref{eq:rate}. All simulations have been carried out in the Julia programming language \cite{juliaLang} using the DifferentialEquations package \cite{diffeqjul}. 

\section{Soliton \label{sec:soliton}}

In a continuous one-dimensional condensate without any driving or dissipation, a family of solitonic solutions exists, that is intimately connected to the integrability of this model, where the dispersion is balanced with a non-linearity and the conserved quantities prohibit the energy from dispersing. Also in systems with driving and dissipation, solitonic solutions exist, the so-called dissipative solitons. They appear when not only dispersion is balanced with a non-linearity, but also the loss by a corresponding gain \cite{akhmediev2005}. 

In this section solitonic excitations in driven-dissipative photon condensates are investigated. In our numerical simulations, we initiate the photon field $\psi(x)$ in the following state corresponding to the soliton in a closed bose gas
\begin{align}\label{sol:eq:psi}
    \psi(x) = \sqrt{n}\bigg[iv+\sqrt{1 -v^2}\tanh{\bigg(\frac{x - x_o}{\xi}\sqrt{1 -v^2}\bigg)}\bigg],
\end{align}
and set the healing length to the healing length \eqref{eq:healinglength} determined in the previous section. This function represents a wavefunction with a phase jump $\Delta\phi = 2 \arccos{(v)}$ dependent on the movement speed of the soliton $v$. A stationary soliton is realized for a phase jump of $\Delta \phi = \pi$. Mean field simulations with our classical field model \eqref{eq:gGP} show that only the stationary on-site soliton is stable. In this dark on site soliton state, the on-site value of $\psi(x)$ in the center of the soliton is exactly zero and therefore the hopping terms on the left and right hand sides exactly cancel each other by destructive interference.

\begin{figure}[h!]
    \centering
    \includegraphics[width = 0.9\linewidth]{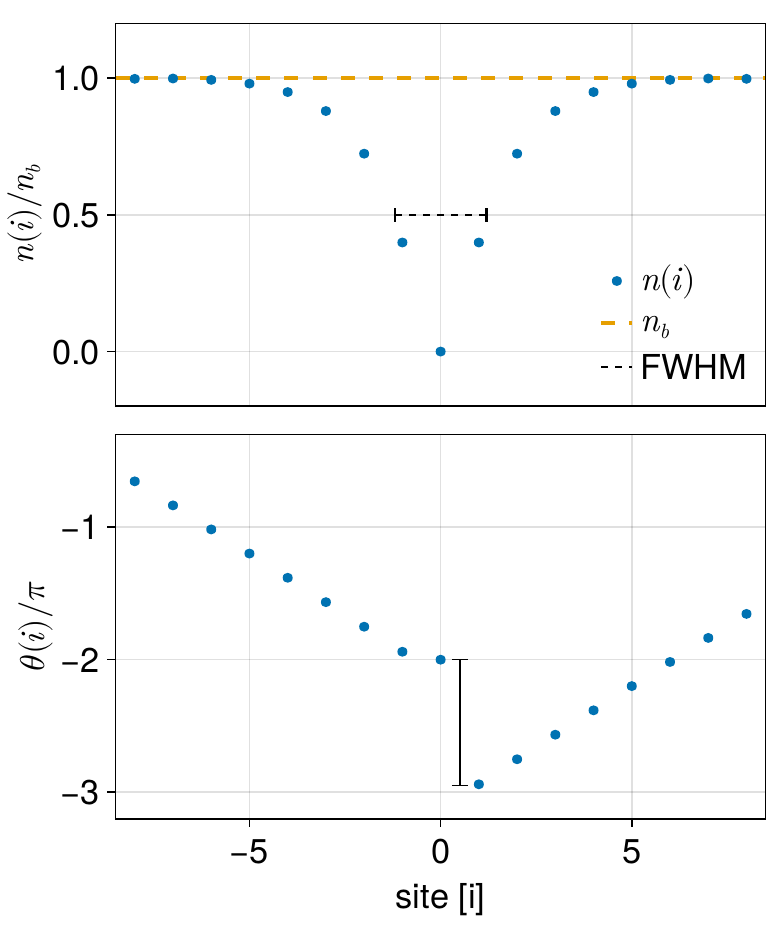}
    \caption{Snapshot of photon density $n$ normalized by the bulk photon value $n_b$ (upper panel) and the phase of the order parameter $\theta$ (lower panel). System parameters: $J=0.02\:meV, \gamma=0.02\:meV, \Delta=-180\:meV, B_{21}=10^{-6}\:meV, n_b=n=5.10^4, M=5.10^{10}, T=25\:meV$ in a $N=1025$ site system.}
    \label{fig:solitonSnapshot}
\end{figure}

Fig. \ref{fig:solitonSnapshot} shows an example of a steady state soliton obtained from the initial condition \eqref{sol:eq:psi}  after a transient where excitations are emitted in both directions. However, we have seen that this soliton solution is unstable with respect to perturbations of the wave function in the center: an infinitesimal perturbation away from $\psi = 0$ for the central value of the wavefunction makes the solution diverge from the solitonic state. Similarly, for stochastic simulations where spontaneous emission is taken into account, we found the soliton to be unstable and therefore conclude that it is not a physically accessible state for a photon condensate.

\begin{figure}[h!]
    \centering
    \includegraphics[width = 1.0\linewidth]{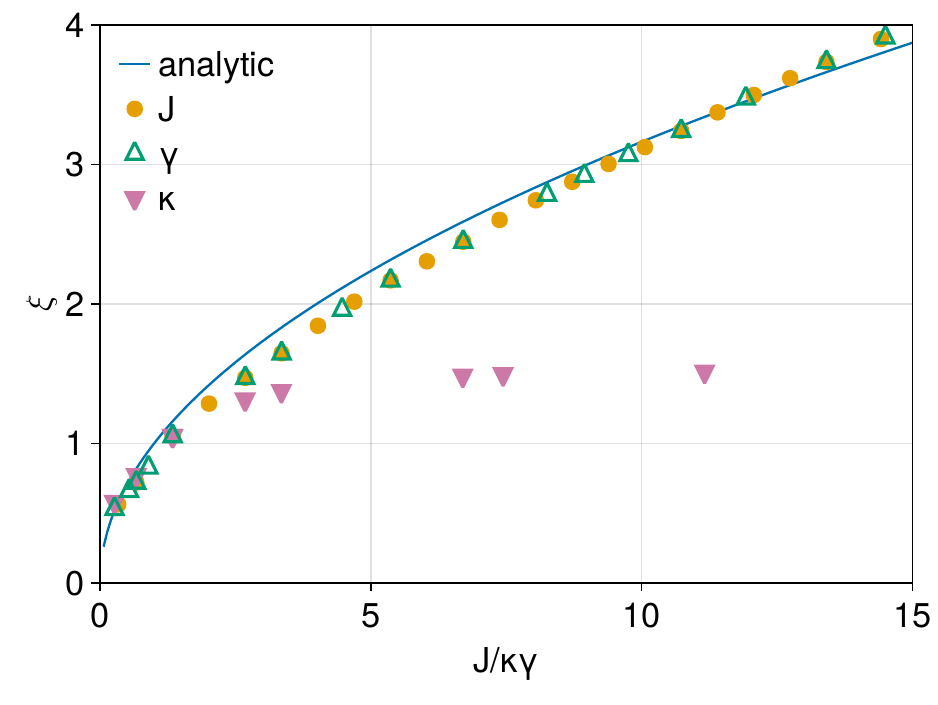}
    \caption{Healing length $\xi$ extracted from the FWHM of a soliton as a function of $J/\kappa\gamma$ by varying the underlying parameters $J$, $\gamma$ and $\kappa$ compared to the analytic healing length. System parameters: $J=0.005 - 0.215\:meV, \gamma=0.00185 - 0.1\:meV, \kappa = 0.09 - 3.73, \Delta=-180\:meV, B_{21}=10^{-6}\:meV, n=5.10^4, M=5.10^{10}, T=25\:meV$ in a $N=1025$ site system.}
    \label{fig:SolitonHealingLength}
\end{figure}

Nevertheless, the mean field soliton can be used to compare its width with the analytically derived healing length \eqref{eq:healinglength}.
For the stationary $v=0$ state of \eqref{sol:eq:psi}, the FWHM of the density satisfies $\xi \approx \ell_{FWHM}/2.49$, which is the same relation as for an equilibrium soliton with a tanh shape.
In figure \ref{fig:SolitonHealingLength}, we show $\xi$ for the mean field photon condensate soliton for various parameter combinations and compare it to the analytical photon condensate healing length \eqref{eq:healinglength}. For most of the simulations, we find an excellent agreement for changing the microscopic parameters $J, \gamma$ and $\kappa$ in the numerics independently.
It should be noted that the parameter $\kappa$ is important for the stability of the soliton and has a more limited parameter range than $J$ and $\gamma$. For small thermalization ($\kappa \le 0.2$) the soliton excitations do not relax under times of $\Delta t J = 10^5$. In Fig. \eqref{eq:healinglength}, this is seen by the much larger deviations between the numerically determined soliton size and the analytical prediction for the filled triangles ($\kappa = 0.09$). For the other parameter values (curves $J$ and $\gamma$), the relaxed shape of the soliton deviates slightly from a hyperbolic tangent. 

\section{Vortices}\label{sec:vortex}

\begin{figure}[h!]
    \centering 
    \includegraphics[angle=270,width = 1.0\linewidth]{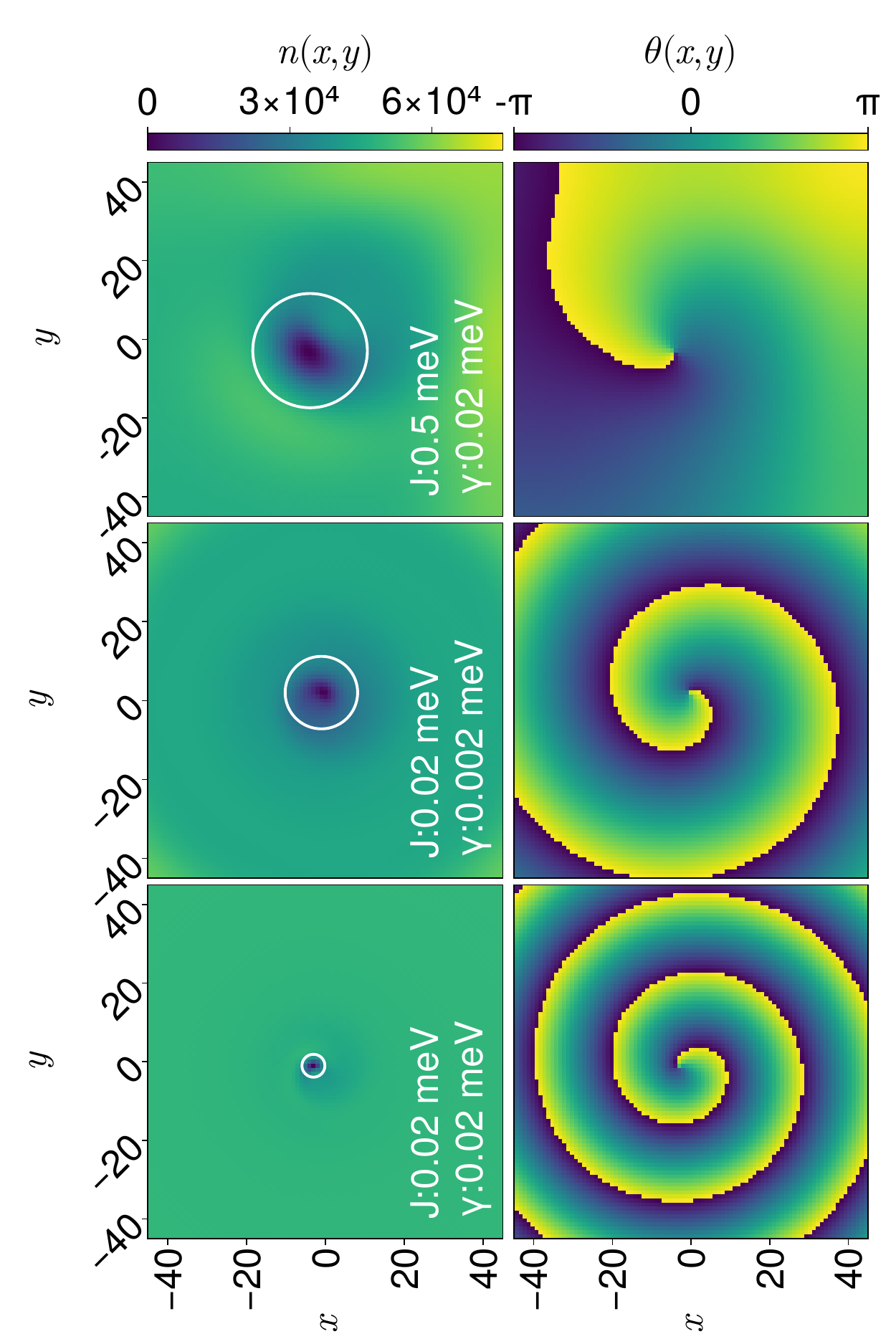}
    \caption{Snapshots of photon density $n$ (upper panels) and the corresponding phase of the order parameter $\theta$ (lower panels). The analytic estimate of the FWHM of the vortex size is shown by the white circle. System parameters: $\Delta =-180\:meV, B_{21}=10^{-6}\:meV, n=5.10^4, M=5.10^{10}, T=25\:meV$ in a $N=129 \times 129$ site system.}
    \label{fig:VortexSnapshot}
\end{figure}

\begin{figure}[h!]
    \centering 
    \includegraphics[width = 1.0\linewidth]{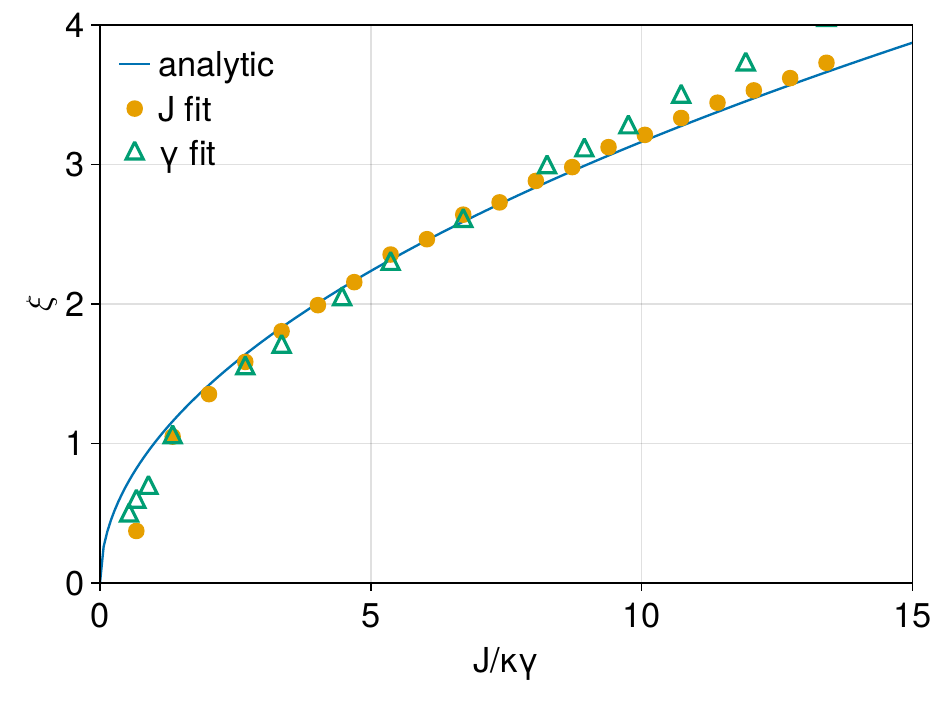}
    \caption{Healing length $\xi$ extracted from the variance of the vortex density ($\alpha\sigma_{x+y}$, $\alpha\approx1.1$) as a function of $J/\kappa\gamma$ by varying the underlying parameters $J$ ($J$ fit curve) and $\gamma$ ($\gamma$ fit curve) compared to the analytic healing length. System parameters: $J=0.01 - 0.2\:meV, \gamma=0.002 - 0.05\:meV, \Delta=-180\:meV, B_{21}=10^{-6}\:meV, n=5.10^4, M=5.10^{10}, T=25\:meV$ in a $N=129 \times 129$ site system.}
    \label{fig:VortexHealingLength}
\end{figure}

While a soliton in 1D is shown to be dynamically unstable, vortex excitations have been shown to be stable in 2D photon condensates \cite{gladilin2021vortex}. 
The size of these vortices is also expected to be of the order of the healing length. Fig. \ref{fig:VortexSnapshot} shows the snapshots of vortices for three different sets of parameters. In contrast to equilibrium condensates, the symmetry of the vortex is broken due to the interplay between azimuthal and radially outward moving currents characteristic of a driven-dissipative superfluid. Fig. \ref{fig:VortexSnapshot} shows that the size of the vortex core is in line with that expected from the healing length (white circles), but also that the shape of the core changes significantly between different parameter sets, complicating the quantitative determination of the core size. The vortex in the right panels ($J=0.5\:meV,\gamma=0.02\:meV$) has a more elongated profile compared to left and center panels. In order to characterize the vortex size, we compute a time averaged total variance in the $x$ and $y$ directions of the photon density, which is summed in the combined standard deviation $\sigma_{x+y} = \sqrt{\sigma_x^2+\sigma_y^2}$ and fitted with a free parameter $\alpha$ to the analytical curve $\xi =\sqrt{J/\kappa\gamma} \approx  \alpha\sigma_{x+y}$. Figure \ref{fig:VortexHealingLength} shows good agreement between analytic and numeric fits, although a deviation between the curves for parameters $J$ and $\gamma$ is observed.  This is attributed to a shape change of the vortex profile due to the underlying parameters, which is not observed for the soliton case.

\begin{figure}[h!]
    \centering 
    \includegraphics[width = 1.0\linewidth]{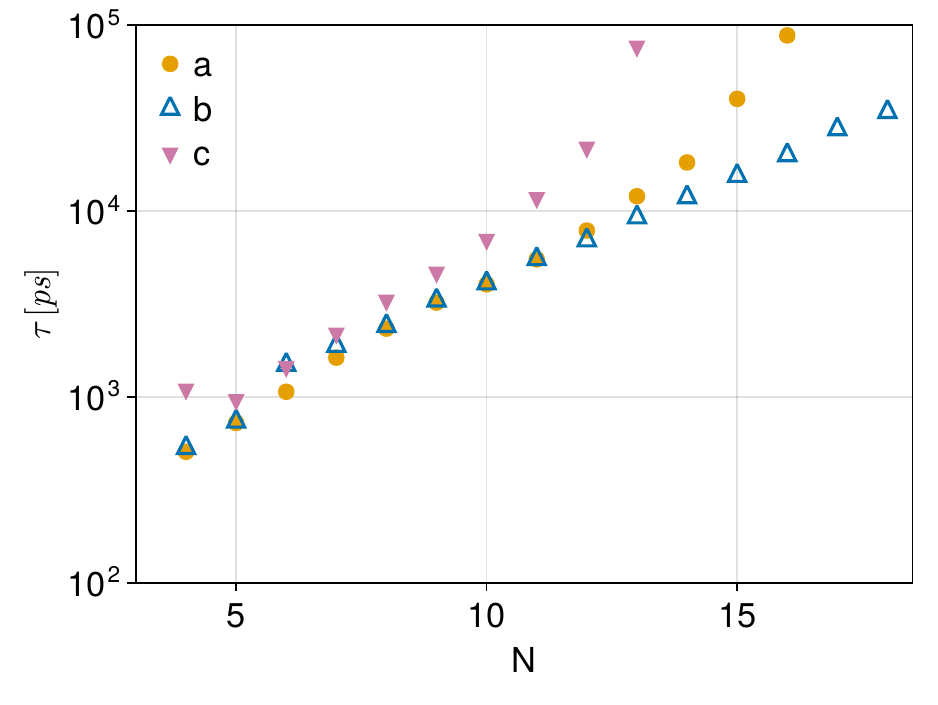}
    \caption{Mean vortex lifetime $\tau$ as a function of the square lattice size $N\times N$. System parameters: $\xi = 1.16, J = 0.02\:meV, \gamma =0.02\:meV, n=5.10^4$ [for data set a], $\xi = 2.32, J = 0.02\:meV, \gamma =0.005\:meV, n=5.10^4$ [for data set b] and $\xi = 1.16, J = 0.02\:meV, \gamma =0.02\:meV, n=8.10^4$ [for data set c], with the other parameters given as $\Delta =-180\:meV, \gamma =0.02\:meV, B_{21}=10^{-6}\:meV, M=5.10^{10},\; T=25\:meV$.}
    \label{fig:VortexLifetime}
\end{figure}

In a finite system, the vortex excitation can disappear by simply leaving the system. The lifetime of a vortex is therefore highly dependent on the size of the system. Fig. \ref{fig:VortexLifetime} shows the effects of increasing the number of cavity sites and changing the vortex parameters on the mean lifetime $\tau$ of a vortex in the lattice. For each datapoint, an initial condition with a vortex in the center is used and the lifetime is determined by averaging over $256$ stochastic runs. All data sets ($a,b$ and $c$) show an exponential growth of the vortex lifetime as a function of number of sites $N$. Up to a system size of $N=11 \times 11$, data sets $a$ and $b$ show similar mean vortex lifetimes, although the healing length differs by a factor of two. From $N=12 \times 12$, the vortex life time in system $a$ increases faster than that in system $b$. Increasing the number of photons improves the lifetime by suppressing the effective noise on the wavefunction, illustrated by comparing system $c$ (higher photon number) with $a$. From a lattice size of $N=5\times5$, all parameter sets show a mean lifetime in the experimentally accessible nanosecond range. 

\begin{figure}[h!]
    \centering 
    \includegraphics[width = 1.0\linewidth]{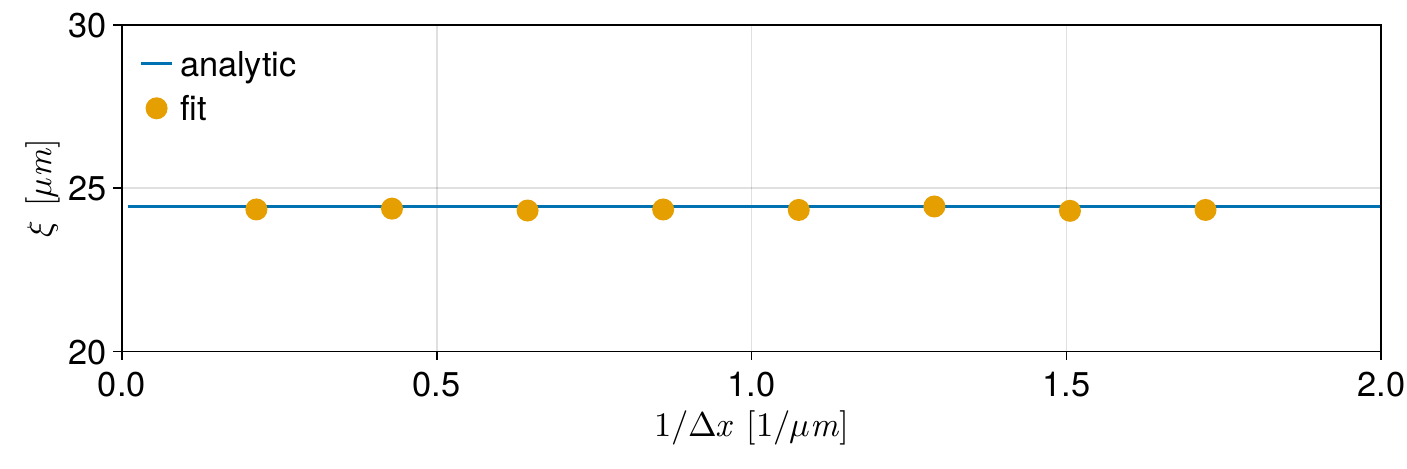}
    \caption{Healing length $\xi$ extracted from the variance of the vortex density ($\alpha\sigma_{x+y}$, $\alpha\approx0.87$) in the continuum limit as a function of inverse grid size $1/\Delta x$ for grids of size $N=128$ to $N=1024$. System parameters: $L = 600\:\mu m,\; \Delta=-140\:meV,\; \gamma=0.02\:meV,\; B_{21}=10^{-5}\:meV,\; n=5.10^4,\; M=10^{8},\; T=25\:meV,\; m=1\: meV \mu m^{2}.$}
    \label{fig:VortexContLimit}
\end{figure}

Finally, we investigate the applicability of the classical field model in the continuum limit. The mean-field continuum equation then reads
\begin{align}
i \frac{\partial \psi(x,t) }{\partial t}  = 
&- (1-i\kappa) \frac{\nabla^2}{2m} \psi(x,t) \nonumber \\
&+
\frac{i}{2}\left[ b_{21}m_2(x,t) - b_{12}
m_1(x,t)-\gamma\right]\psi(x,t) \nonumber,
 \label{eq:gGP_c}
\end{align}
where $m_{1}(x,t)$ and $m_2(x,t)$ denote the densities of unexcited and excited molecules, respectively. The continuous Einstein coefficients are given by $b_{12}$ and $b_{21}$. These definitions result in rates that are independent of the grid size because $b_{12}m_1(x,t) = B_{12}M_1(x,t)$ and $b_{21}m_2(x,t) = B_{21}M_2(x,t)$. 
In our numerical simulations, we of course have to discretize the continuous space. Representing the second derivative by the finite difference expression, the grid size then corresponds to a hopping rate $J = 1/(2 m (\Delta x)^2)$. We probe grid sizes between $N=128$ and $N=1024$ with system length $L = 600\: \mu m$, resulting in a coupling range of $J \approx  0.2\: - 13\:meV$. Fig. \ref{fig:VortexContLimit} shows the size of the vortex as a function of a denser grid by increasing the number of grid points $N$. From $1/\Delta x = 0.21\:1/\mu m$, corresponding to $N=128$, the numerics are converged and stable over the entire range of probed grid sizes. This convergence shows that our model is applicable to the continuum limit as well. 
The estimated vortex core size for our realistic parameters should be within the range of experimentally achievable system sizes.

\section{Conclusions and Outlook}\label{sec:conclusion}

We presented an analytic derivation for the healing length of a driven-dissipative dye-based photon condensate by considering small perturbations and showed that this gives a very good estimate for both solitons in one dimensional condensates and vortices in two dimensions.
We show that stationary solitons are dynamically unstable under noise and only in the mean-field limit do on-site solitons thermalize for a sufficiently large thermalization parameter $\kappa$. 
In contrast to solitons, vortices have been shown to be robust to noise and demonstrate non trivial motion \cite{gladilin2021vortex}. We further explored the stability of the vortex parameter regime and used the core size to test our expression for the healing length. The dependence of the vortex shape on system parameters slightly changes the healing length compared to the length extracted from the soliton, but still closely agrees with the analytic formula. Additionally, we show that for a small thermalization parameter, the solitons and vortices do not relax properly. We also analyzed the mean lifetime of vortices as a function of lattice size. All probed systems show a roughly exponential growth of the vortex lifetime when increasing the system size. Already for lattices as small as $5\times5$, we found vortex lifetimes in the experimentally accessible $ns$ range. Finally, we have investigated the applicability of the classical field model \cite{gladilin2020classical} to the mean-field continuum limit, which shows excellent convergence for grid sizes between $\approx [0.2-1.75]\: 1/\mu m$. 
In this work we have assumed a perfectly homogeneous lattice and neglected disorder. Imperfections in the lattice as well as the study of how the vortices can be experimentally initialized will be interesting topics of further investigation.

\section{Acknowledgements}
We thank Julian Schmitt and Martin Weitz for valuable discussions on photon condensation and vortices. MY is financially supported through the FWO-Vlaanderen through research grant G0B2923N.
\bibliography{biblio.bib}
\end{document}